\documentclass{article}

\usepackage[preprint]{neurips_2026}

\usepackage[utf8]{inputenc} 
\usepackage[T1]{fontenc}    
\usepackage{hyperref}       
\usepackage{url}            
\usepackage{booktabs}       
\usepackage{amsfonts}       
\usepackage{nicefrac}       
\usepackage{microtype}      
\usepackage{xcolor}         
\usepackage{graphicx}
\usepackage{float}
\usepackage{amsmath, amssymb}

\title{Arbitrage Analysis in Polymarket NBA Markets}

\author{%
Guang Cheng\thanks{Corresponding author: Prof. Guang Cheng (\texttt{guangcheng@ucla.edu})}, Jiaxin Yang and Haoxuan Zou\\
  University of California, Los Angeles\\
  Los Angeles, CA 90095 \\
  \today
}

\begin{document}

\maketitle

\begin{abstract}
While decentralized prediction markets like Polymarket have gained significant traction, their market microstructure and high-frequency pricing efficiency remain underexplored. This paper conducts a systematic empirical analysis of algorithmic arbitrage within Polymarket's NBA game markets. By reconstructing continuous market states from over 75 million limit order book snapshots across 173 games, we evaluate the frequency, duration, and profitability of both single-market and combinatorial arbitrage opportunities. Our findings demonstrate profound microstructural efficiency. Single-market anomalies are exceedingly rare, yielding only 7 executable in-game episodes that persist for a median duration of just 3.6 seconds. Combinatorial inefficiencies are more frequent, producing 290 active episodes overwhelmingly concentrated in the final minutes of live play. While combinatorial execution yields a statistically meaningful median return of 101 basis points, we find that the theoretical "Middle" jackpot is never empirically realized. Furthermore, execution is severely bottlenecked by shallow order book depth, with 76.9\% of combinatorial opportunities constrained to an average executable size of just 14.8 shares. Ultimately, while executable mispricings exist, they are structurally bounded by liquidity, confining risk-free extraction strictly to the retail scale.
\end{abstract}

\section{Introduction}
\label{sec:intro}

Prediction markets aggregate dispersed information to forecast future events, with prices reflecting the market's consensus probability. Polymarket has recently emerged as a prominent decentralized prediction market on the Polygon blockchain, attracting substantial liquidity across diverse domains. While academic interest in Polymarket has grown, existing research predominantly focuses on its efficacy in political forecasting \cite{ng2025election}. Comparatively little attention has been paid to the market's microstructure and trading behavior, largely due to the data engineering challenges of decoding complex on-chain data.

Polymarket operates a hybrid-decentralized Central Limit Order Book (CLOB) that facilitates high-speed, off-chain order matching. By capturing continuous, high-resolution snapshots of this limit order book, we gain a precise, point-in-time view of market depth and liquidity. This presents a unique opportunity to systematically analyze market inefficiencies as they emerge in real time. This paper exploits these high-frequency order book snapshots to conduct an empirical study of algorithmic arbitrage in Polymarket. Specifically, we quantify two types of arbitrage opportunities: (1) single-market arbitrage arising from mispricings across complementary outcome shares within a single market, and (2) combinatorial arbitrage that spans multiple related markets simultaneously.

\begin{figure}[h]
    \centering
    \includegraphics[width=0.8\textwidth]{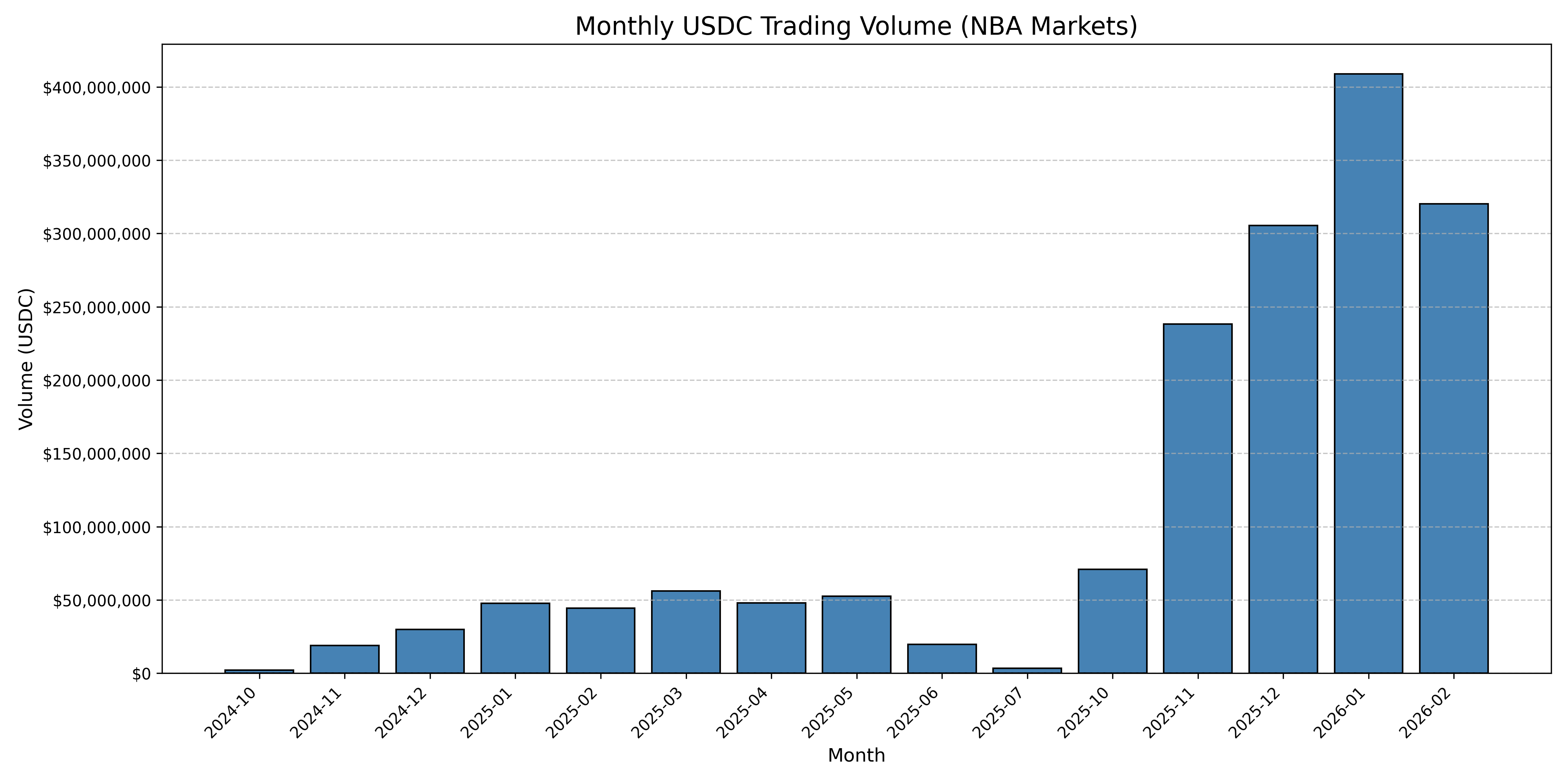}
    \caption{Monthly trading volume of NBA markets from Oct 2024 to Feb 2026}
    \label{fig:monthly_nba_volume}
\end{figure}

We select NBA game markets due to their high liquidity, structural regularity, and frequency. NBA markets account for nearly 30\% of Polymarket’s sports activity in 2025, with trading volume increasing from \$51M in 2024 to \$0.89B in 2025 (Figure \ref{fig:monthly_nba_volume}), representing approximately a 17.45 times growth. Furthermore, the regular season's 1,230 games provide a vast, homogeneous sample of markets with consistent resolution logic, offering the statistical power necessary to rigorously evaluate high-frequency arbitrage opportunities.

In platforms where mutually exclusive outcome probabilities must sum to one, mispricing creates risk-free profit opportunities. While prior work by \cite{saguillo2025arbitrage} empirically analyzes Polymarket arbitrage using {\em executed trade data}, our work introduces a key methodological advancement by leveraging {\em high-resolution order book snapshots}. This approach allows us to detect arbitrage opportunities in real time as they emerge, measure their persistence, and determine their actual executability against prevailing liquidity, offering a more precise characterization of market efficiency. Critically, executability proves to be the binding constraint: shallow order book depth structurally caps the scale of risk-free extraction, an empirical manifestation of the limits-to-arbitrage frictions formalized by 
\citet{shleifer1997limits}. This liquidity-bounded pattern has been documented on Polymarket in the context of political event contracts 
\citep{tsang2026anatomy}; we show it holds equally, and perhaps more severely, in sports markets where contracts resolve within hours and liquidity has less time to accumulate.

\section{Background and Market Structure}

\subsection{Polymarket Mechanics and Pricing}
Polymarket is a decentralized prediction market operating on the Polygon blockchain via a peer-to-peer exchange model. Markets are grouped into discrete ``events'' (e.g., an individual NBA game). Each market poses a specific, typically binary question, traded via ``Yes'' and ``No'' outcome tokens (ERC-1155). These tokens are fully collateralized: every complementary Yes/No pair is backed by exactly \$1.00 of USDC locked in a smart contract. 

Token prices strictly range between \$0.00 and \$1.00, functioning as the market's collective probability estimate for a given outcome. Because outcomes are mutually exclusive, the combined price of the ``Yes'' and ``No'' tokens for a given market should naturally sum to \$1.00, discounting bid-ask spreads. When prices deviate from this bounded sum, single-market arbitrage opportunities arise; see Section 3. 

Trading is facilitated through CLOB. Orders are submitted and matched by an off-chain operator to ensure low latency and zero gas fees, while the atomic swap of tokens and USDC is settled on-chain. 

\subsection{NBA Market Typology and Structural Dependencies}
Every NBA game generates a cluster of structurally related, game-level markets. These primarily consist of:
\begin{itemize}
    \item \textbf{Moneyline:} A binary market predicting the outright winner.
    \item \textbf{Spreads:} Markets predicting the outcome based on specific point differentials.
    \item \textbf{Totals (Over/Under):} Markets predicting the combined score of both teams.
    \item \textbf{Player Props:} Markets predicting the statistical performance of individual players, such as points, rebounds, or assists.
\end{itemize}

A single NBA game event can encompass around 45 distinct markets, with about 14 Moneyline/Spread/Totals markets and over 30 player proposition markets. As shown in Table \ref{tab:volume_by_market_type}, these markets attract highly variable liquidity and volume. 

\begin{table}[htbp]
    \centering
    \footnotesize
    \begin{tabular}{lrrrr}
    \toprule
    \textbf{Market Type} & \textbf{No. of Markets} & \textbf{Mean Trades} & \textbf{Total Volume (USDC)} & \textbf{Mean Volume} \\
    \midrule
    Full Moneyline  & 176   & 10,412.13 & 251,207,897.58 & 1,427,317.60 \\
    Full Spread     & 769   &    366.78 &  41,057,412.35 &    53,390.65 \\
    Full Total      & 1,162 &    185.60 &  21,804,680.06 &    18,764.78 \\
    Player Prop     & 5,074 &     13.32 &   1,285,140.84 &       253.28 \\
    1H Moneyline    & 172   &     16.01 &     161,774.96 &       940.55 \\
    1H Total        & 457   &      6.12 &      48,107.17 &       105.27 \\
    1H Spread       & 341   &      5.94 &      26,928.93 &        78.97 \\
    \bottomrule
    \end{tabular}
    \caption{Summary of trading volume across NBA market types.}
    \label{tab:volume_by_market_type}
\end{table}

Crucially, while these markets resolve based on correlated real-world outcomes (e.g., a team covering a large spread is mathematically guaranteed to win the moneyline), they trade on completely isolated order books. Polymarket's smart contracts do not natively cross-margin or dynamically link the pricing of these separate markets. This structural isolation frequently leads to transient pricing misalignments across dependent markets, forming the mathematical basis for the combinatorial arbitrage strategies analyzed in Section 4.

\subsection{Data Architecture and Constraints}
To evaluate these arbitrage opportunities, this study captures real-time market depth rather than relying on executed trade histories. We utilize Polymarket's off-chain CLOB API (\texttt{GET /book} endpoint) to construct point-in-time snapshots of the limit order book. While the API provides deeper market depth, our analysis strictly isolates and evaluates the top-of-book Best Bid and Best Ask (Level 1 data) to ensure we are measuring the most conservative, highly executable liquidity available.

This infrastructure introduces a specific methodological constraint: fetching the complete state across all active NBA markets requires a polling cycle of 3.6 to 5.5 seconds. Consequently, our dataset is bounded by this polling frequency, which restricts the observation of transient anomalies. Because market states are observed discretely, episode duration is estimated using a forward-looking difference. If an arbitrage is detected at snapshot $n$, the episode is credited with the time gap until the subsequent snapshot at $n+1$. This discrete sampling means duration measurements are structural estimates rather than exact lifespans. For any given recorded episode, this calculation may overestimate its actual duration if the inefficiency is corrected by market makers immediately after snapshot $n$ is recorded. Conversely, on a macro level, this polling cadence systematically underestimates the total frequency of market inefficiencies, as flash arbitrages that open and close entirely within a single polling gap evade detection completely. Nevertheless, this cadence establishes a rigorous lower bound on market inefficiency; any opportunity that survives a full multi-second polling cycle represents a highly actionable structural dislocation, ensuring our empirical findings capture genuine, exploitable inefficiencies rather than unexecutable micro-second noise.

\subsection{Mirrored Orderbook}
A unique feature of Polymarket CLOB is its mirrored liquidity design. There is only one shared pool of underlying liquidity for any binary market. The matching engine automatically reflects orders across the complementary tokens. For example, if a market maker places a limit order to buy 100 shares of Outcome A (YES) at \$0.40, the off-chain engine simultaneously generates a synthetic limit order to sell 100 shares of Outcome B (NO) at \$0.60. The inversion is a purely mathematical reflection ($\$1 - P_{original}$). Consequently, querying the CLOB API for Outcome B yields an effective orderbook that combines direct orders placed on Outcome B with the synthetic, mirrored orders originating from Outcome A.

The mirrored nature of the liquidity pool introduces a systemic risk of double-counting arbitrage signals. If a severe mispricing occurs, an anomaly where $Ask_A + Ask_B < 1.00$ will automatically manifest as a crossed synthetic book where $Bid_A + Bid_B > 1.00$. Naively summing the profit from both the Long and Short signals would falsely double the executable profitability of the market.

To resolve this, the analysis pipeline evaluates both paths independently at every timestamp available in the data to capture all possible price discrepancies, but enforces a strict deduplication protocol. The analysis consolidates the arbitrages by isolating the simultaneous signals and retains only one path in the results.

\section{Single Market Arbitrage}

This section evaluates the frequency, duration, and profitability of single-market arbitrage in Polymarket NBA events. By analyzing over 75 million limit order book snapshots, we demonstrate that the market exhibits profound microstructural efficiency. We identify only 7 valid, exploitable in-game arbitrage episodes across 3,042 markets. When inefficiencies did occur, they were localized entirely within the live-game phase and corrected swiftly by automated participants, persisting for a median duration of just 3.6 seconds, all below the mechanical polling threshold of the data collection pipeline. Furthermore, we find that spread markets experienced deeper liquidity vacuums during these transient mispricings compared to moneyline markets.

\subsection{Dataset and Methodology Summary}
The analysis utilized high-frequency Limit Order Book (LOB) snapshots collected from the Polygon blockchain for 173 NBA games between February 4, 2026, and March 4, 2026. The dataset comprised 75,088,497 LOB snapshots. 

In a strictly binary prediction market, the underlying asset resolves to either True or False, guaranteeing a combined payout of exactly \$1.00 for holding one share of each outcome. A risk-free arbitrage opportunity exists whenever the cost to acquire exposure to all mutually exclusive outcomes is misaligned with this payout. We evaluated two execution paths defined in \cite{saguillo2025arbitrage}:
\begin{enumerate}
    \item \textbf{The Buy Path (Long Arbitrage):} Simultaneously buying all outcomes from the ask side for a combined cost strictly less than \$1.00:
    $$\text{Ask}_{A} + \text{Ask}_{B} < 1.00$$
    \item \textbf{The Mint-and-Sell Path (Short Arbitrage):} Locking \$1.00 of USDC collateral into the smart contract to mint shares of both outcomes, and immediately selling them to the bid side for combined revenue strictly greater than \$1.00:
    $$\text{Bid}_{A} + \text{Bid}_{B} > 1.00$$
\end{enumerate}

To ensure the identified inefficiencies represented actionable trading opportunities rather than phantom signals, strict execution constraints were applied. We exclusively evaluated top-of-book (Level 1) data, enforced a minimum executable liquidity bottleneck of \$10 USDC, and strictly deduplicated signals to account for Polymarket's mirrored liquidity pool. Crucially, we excluded all order book snapshots recorded after the physical conclusion of the sporting event. Comprehensive details regarding state alignment, network latency controls, and duration measurement are provided in Appendix \ref{sec:appendix_single_market}.

\subsection{Empirical Results}

\textbf{Validation of Post-Game Episode Exclusion:}
The raw detection pipeline initially identified 37 theoretical arbitrage episodes, but 30 of these (81.1\%) occurred strictly in the post-game phase. On Polymarket, a mechanical delay exists between the physical conclusion of a game and the final oracle resolution. During this window, market makers withdraw active quotes, severely hollowing out the order book. As Table \ref{tab:arb_exclusion_spread} demonstrates, the median bid-ask spread exploded to 7,532.65 bps post-game. Stale limit orders may appear to cross, but there is no active liquidity available to execute. Therefore, excluding these post-game artifacts was mathematically necessary, leaving 7 valid, executable in-game episodes.

\begin{table}[htbp]
    \centering
    \caption{Order book liquidity comparison by game phase.}
    \label{tab:arb_exclusion_spread}
    \begin{tabular}{lc}
        \toprule
        \textbf{Market Phase} & \textbf{Median Bid-Ask Spread (bps)} \\
        \midrule
        Pre-Game & 392.20 \\
        In-Game & 1030.90 \\
        Post-Game & 7532.65 \\
        \bottomrule
    \end{tabular}
\end{table}

\textbf{Overall Market Results:}
Despite processing over 75 million snapshots, the analysis identified only 7 
valid arbitrage episodes, indicating that the vast majority of markets exhibit 
perfect microstructural efficiency. True arbitrage activity was exclusively an 
in-game phenomenon (Table \ref{tab:phase_summary}). Assuming a maximum capital deployment of \$100 per episode, the aggregate capped profit totaled \$210.19. 
However, this figure is heavily right-skewed by a single outlier game in which 
a late blowout caused market makers to withdraw their quotes, leaving stale 
mispriced orders on the book that briefly crossed before automated participants 
restored efficiency. Stripping this outlier, the median yield is a more representative 11.0\% (\$11.01 per episode). Assuming unlimited liquidity, the theoretical uncapped profit was \$4,418.44, illustrating the gap between theoretical inefficiencies and practically extractable value.

\begin{table}[htbp]
\centering
\caption{Summary of valid arbitrage episodes and profitability by game phase.}
\label{tab:phase_summary}
\resizebox{\columnwidth}{!}{%
\begin{tabular}{lccccc}
\toprule
\textbf{Game Phase} & \textbf{Episodes} & \textbf{\% Time in Arb} & \textbf{Median Dur.\ (s)} & \textbf{Capped Profit} & \textbf{Uncapped Profit} \\
\midrule
Pre-Game & 0  & 0.0000\% & --- & \$0.00   & \$0.00   \\
In-Game  & 7  & 0.0001\%  & 3.614 & \$210.19  & \$4,418.44 \\
\midrule
\textbf{Total} & \textbf{7} & --- & \textbf{3.614} & \textbf{\$210.19} & \textbf{\$4,418.44} \\
\bottomrule
\end{tabular}%
}
\end{table}

\begin{figure}[H]
\centering
\includegraphics[width=0.8\textwidth]{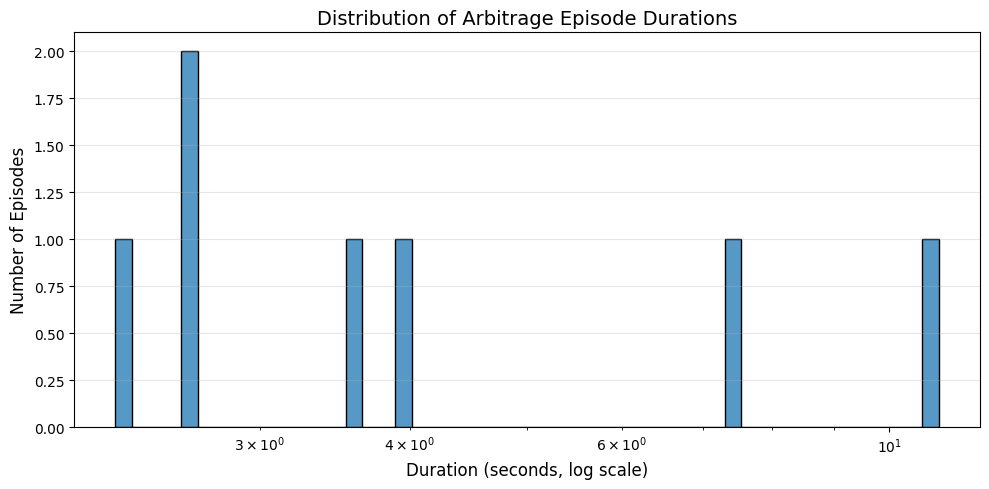}
\caption{Distribution of arbitrage episode durations.}
\label{fig:duration_dist}
\end{figure}

\textbf{Speed of Arbitrage and Market Resilience:}
When true pricing inefficiencies did occur, they were corrected swiftly. The median duration of an arbitrage episode was 3.614 seconds (Figure \ref{fig:duration_dist}). This distribution aligns closely with the mechanical reality of our 3.6 to 5.5 second API polling cycle. The absence of sub-second artifacts confirms these episodes represent genuine mispricings that survived at least one full network polling sweep, indicating automated market makers occasionally require a few seconds to fully reprice an orderbook following a sudden live-game shock.

\textbf{Profitability by Market Type:}
While occurrences were equally rare (3 episodes each for Moneyline and Spread), Spread markets generated disproportionately higher theoretical profit (Table \ref{tab:market_type_summary}). The 3 Spread episodes produced \$194.08 in capped profit compared to just \$5.10 across the 3 Moneyline episodes. This suggests that while Moneyline markets may break briefly, their liquidity remains dense. Conversely, Spread markets experience deeper, albeit brief, liquidity vacuums. Notably, in 85.7\% of the valid episodes, resting liquidity was sufficient to fully absorb a retail-sized (\$100) execution without bottlenecking.

\begin{table}[htbp]
\centering
\caption{Arbitrage episodes and profitability by market type.}
\label{tab:market_type_summary}
\resizebox{\columnwidth}{!}{%
\begin{tabular}{lcccc}
\toprule
\textbf{Market Type} & \textbf{Episodes} & \textbf{Median Duration (s)} & \textbf{Capped Profit} & \textbf{Uncapped Profit} \\
\midrule
Moneyline & 3  & 3.964 & \$5.10  & \$39.20  \\
Spread  & 3  & 2.652 & \$194.08 & \$4,368.23  \\
Total (Points) & 1 & 2.276 & \$11.01 & \$11.01  \\
\midrule
\textbf{Overall} & \textbf{7} & \textbf{3.614} & \textbf{\$210.19} & \textbf{\$4,418.44} \\
\bottomrule
\end{tabular}%
}
\end{table}

\begin{figure}[H]
\centering
\includegraphics[width=0.8\textwidth]{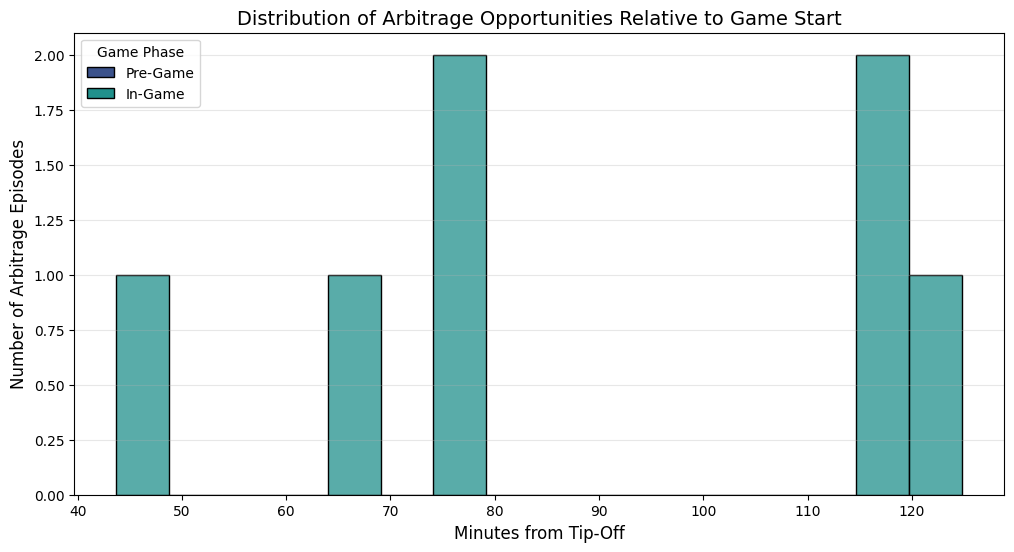}
\caption{Distribution of arbitrage opportunities relative to game start.}
\label{fig:RQ1_arb_dist}
\end{figure}

\textbf{Temporal Distribution Relative to Game Start:}
As established, 100\% of the 7 valid episodes occurred during live play. However, due to the limited number of arbitrage events, no significant temporal pattern was observed relative to specific periods within the game time (Figure \ref{fig:RQ1_arb_dist}).

\section{Combinatorial Arbitrage}
\label{sec:combinatorial}

This section evaluates the frequency and profitability of combinatorial arbitrage between NBA Moneyline and Point Spread markets. Analyzing 8.59 million combinatorial market states revealed 290 active, executable arbitrage episodes, overwhelmingly concentrated during the final minutes of live play. While the strategy yielded an economically meaningful median return of 101 basis points per execution, the theoretical "Middle" jackpot was never realized empirically. Furthermore, execution was severely bottlenecked by shallow order book depth, confining the strategy to retail-sized capital deployment.

\subsection{Problem Setting and Methodology Summary}
\label{sec:problem_setting}

We define two binary derivative contracts for the Favorite team ($A$) against the Underdog team ($B$). Let $\Delta = S_A - S_B$ represent the final point differential. In the NBA, games cannot end in a tie, implying $\Delta \in \mathbb{Z} \setminus \{0\}$.
\begin{itemize}
    \item \textbf{Moneyline ($ML_A$):} Pays if Team A wins outright ($\mathbb{I}_{\{\Delta \ge 1\}}$).
    \item \textbf{Spread ($Sp_A$):} Pays if Team A wins by a margin strictly greater than handicap $h$ ($\mathbb{I}_{\{\Delta > h\}}$, where $h \ge 1$).
\end{itemize}

Because $\{\Delta > h\} \subset \{\Delta \ge 1\}$, the Spread outcome is a strict mathematical subset of the Moneyline. In an efficient market, the Spread contract's price must be less than or equal to the Moneyline contract's price. A theoretical arbitrage exists if the market misprices this relationship:
$$\text{Bid}(Sp_A) > \text{Ask}(ML_A)$$

Because Polymarket does not support direct short selling, traders cannot sell the overpriced $Sp_A$. Instead, they must construct a \textit{synthetic short} by purchasing the complementary token, $Sp_B$ (where $\Delta < h$). The risk-free execution condition requires the combined ask prices to be strictly less than the \$1.00 guaranteed payout:
$$\text{Ask}(ML_A) + \text{Ask}(Sp_B) < 1.00$$

\textbf{The ``Middle'' Jackpot:} This combinatorial strategy creates an asymmetric upside. If the final margin falls precisely in the gap between the Moneyline and the handicap (e.g., $\Delta = 1$ for an $h=1.5$ spread), both contracts resolve to ``Yes.'' This yields a payout of \$2.00, doubling the return on risk-free capital (Table \ref{tab:payoff_matrix}).

\begin{table}[h]
    \centering
    \caption{Payoff Matrix for Synthetic Combinatorial Arbitrage (Team A ML + Team B +1.5)}
    \label{tab:payoff_matrix}
    \begin{tabular}{lcccc}
        \toprule
        \textbf{Game Scenario} & \textbf{$\Delta_{A}$} & \textbf{Team A ML} & \textbf{Team B +1.5} & \textbf{Total} \\
        \midrule
        Team A Wins Big    & $\ge 2$ & \$1.00 & \$0.00 & \$1.00 \\
        Team A Wins Narrow & $= 1$   & \$1.00 & \$1.00 & \textbf{\$2.00} \\
        Team B Wins        & $\le 0$ & \$0.00 & \$1.00 & \$1.00 \\
        \bottomrule
    \end{tabular}
\end{table}

\textbf{Methodology Summary:} Detecting combinatorial arbitrage between isolated markets requires strict synchronization. We implemented a Unified State Machine to merge asynchronous Moneyline and Spread data streams into a single chronological timeline. Missing states were imputed using a strict forward-fill operation to prevent look-ahead bias, and episode durations were capped to control for exchange outages. Complete data engineering details are provided in Appendix \ref{sec:appendix_combinatorial}.

\subsection{Empirical Results}

\textbf{Market Coverage and Frequency:}
The analysis pipeline processed data from the NBA events between February 4, 2026, and March 4, 2026. We evaluated 8.59 million discrete combinatorial market states, detecting 523 candidate arbitrage episodes. After excluding 233 un-executable post-game artifacts, the final dataset yielded 290 active, executable episodes (a median of 2.00 episodes per game).

As shown in Table \ref{tab:coverage}, market inefficiencies were overwhelmingly concentrated during live play. The Pre-Game phase yielded only 11 episodes, a trivial 0.0034\% time-in-arbitrage rate. The In-Game phase generated 279 active episodes, indicating that rapid shifts in win probability driven by live scoring create cross-market dislocations. However, the order book corrected these swiftly, ensuring the market spent the vast majority of its time in an efficient state.

\begin{table}[h]
    \centering
    \caption{Market coverage and frequency of arbitrage by game phase.}
    \label{tab:coverage}
    \resizebox{\columnwidth}{!}{%
    \begin{tabular}{lcccc}
        \toprule
        \textbf{Game Phase} & \textbf{Evaluated States} & \textbf{Total Arb Episodes} & \textbf{\% of Time in Arb State} & \textbf{Median Duration (s)} \\
        \midrule
        Pre-Game & 6,756,454 & 11  & 0.0034\% & 11.0 \\
        In-Game  & 1,834,885 & 279 & 0.1762\% & 16.0 \\
        \bottomrule
    \end{tabular}%
    }
\end{table}

\begin{figure}[H]
    \centering
    \includegraphics[width=0.8\textwidth]{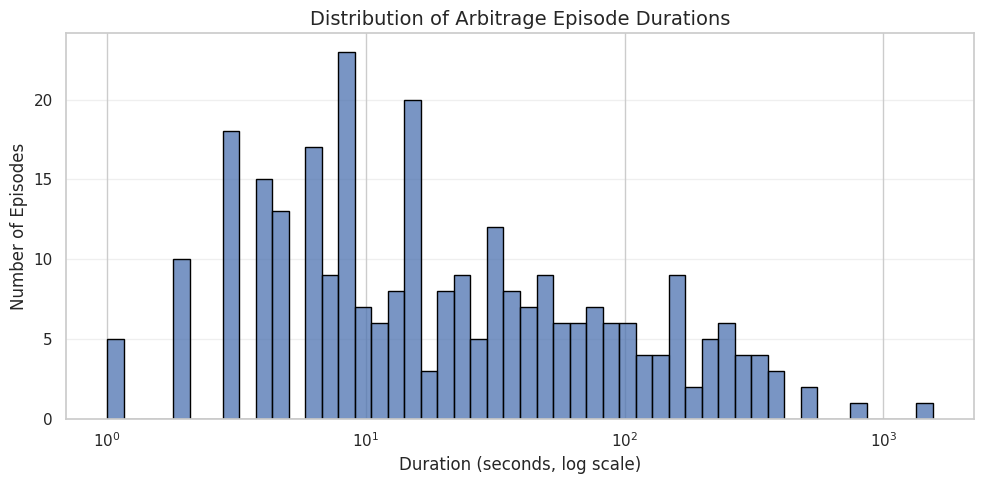}
    \caption{Distribution of arbitrage episode durations.}
    \label{fig:RQ2_arb_duration}
\end{figure}

\begin{figure}[H]
    \centering
    \includegraphics[width=0.8\textwidth]{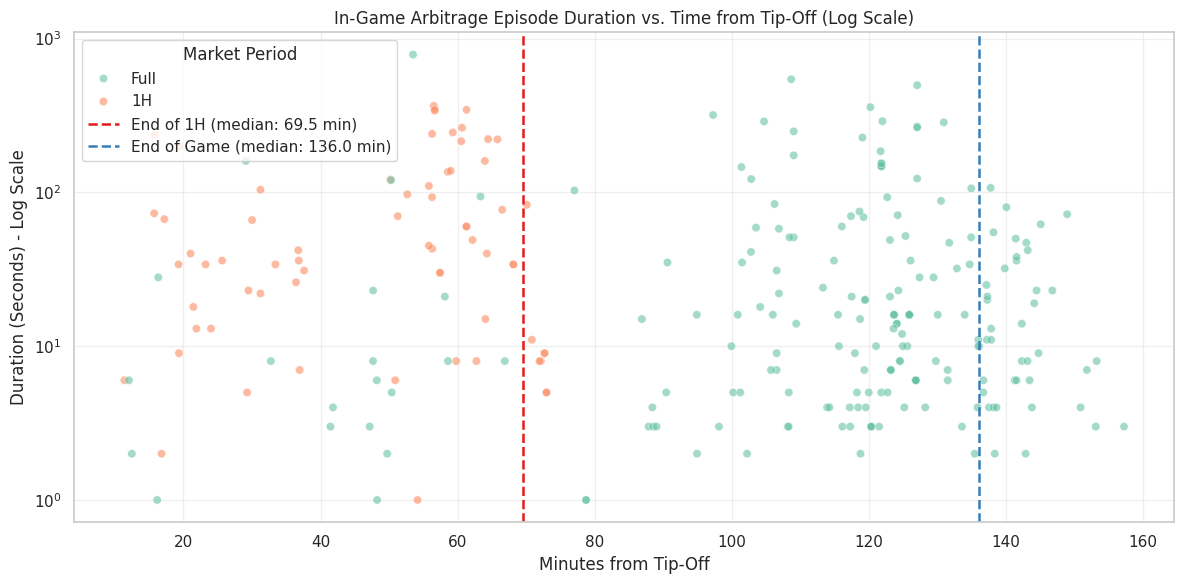}
    \caption{In-Game arbitrage duration vs. time from tip-off.}
    \label{fig:RQ2_market_breakdown_timeline}
\end{figure}

Arbitrage activity was heavily right-skewed and concentrated in the final stretch of the game (Figure \ref{fig:RQ2_market_breakdown_timeline}). End-game scenarios, where winning probabilities approach terminal values and small scoring events cause abrupt shifts in implied odds, were the primary drivers of these dislocations. These opportunities were highly transient (Figure \ref{fig:RQ2_arb_duration}). The median episode lasted 16 seconds, and 17.2\% of all active episodes lasted 4.0 seconds or less, narrowly surviving the baseline polling cycle of the data collection pipeline.


\textbf{Absence of Jackpot Realizations:}
An ex-post resolution analysis was conducted on all candidate episodes to determine if the theoretical ``Middle'' payout materialized. Empirically, zero episodes resulted in a jackpot realization; every combinatorial pair resolved to the baseline \$1.00 payout. Capturing the double payout requires the simultaneous occurrence of a transient cross-market pricing dislocation and a terminal point differential that falls precisely within the handicap gap. This intersection proved exceedingly unlikely over the sample period. Consequently, algorithmic participants cannot assign meaningful probability weight to this incremental upside in rigorous expected-value calculations.

\textbf{Profitability and Execution Bottlenecks:}
We assessed economic viability using a "One-Shot" execution paradigm capped at a \$100 theoretical budget per episode (Table \ref{tab:profitability}). Across 290 active episodes, the aggregate capped profit was \$559.59. While the median yield of 101.01 basis points is economically meaningful in percentage terms, absolute returns were heavily constrained by market depth. 

\begin{table}[h]
    \centering
    \caption{Aggregate and Median Profitability by Game Phase.}
    \label{tab:profitability}
    \begin{tabular}{lcccc}
        \toprule
        \textbf{Phase} & \textbf{Episodes} & \textbf{Median Yield (bps)} & \textbf{Capped Profit} & \textbf{Uncapped Profit} \\
        \midrule
        In-Game  & 279 & 101.01 & \$527.17 & \$1,924.79 \\
        Pre-Game & 11  & 309.28 & \$32.42  & \$107.96   \\
        \midrule
        \textbf{Total} & \textbf{290} & \textbf{101.01}\textsuperscript{*} & \textbf{\$559.59} & \textbf{\$2,032.75} \\
        \bottomrule
        \multicolumn{5}{p{0.9\textwidth}}{\vspace{2pt}\footnotesize{\textsuperscript{*}\textit{Note:} The overall median yield is identical to the In-Game median because In-Game episodes comprise 96\% of the total dataset.}} \\
    \end{tabular}
\end{table}

The discrepancy between the uncapped theoretical profit (\$2,032.75) and the capped figure (\$559.59) directly quantifies this liquidity bottleneck. In 76.9\% of all episodes, the \$100 budget could not be fully deployed, with the average executable size restricted to approximately 14.79 synthetic shares. This structural constraint echoes the limits-to-arbitrage mechanism described by \citet{shleifer1997limits}: even when a mispricing is unambiguously identified, execution and funding frictions prevent its full correction. \citet{tsang2026anatomy} document a closely analogous pattern in Polymarket's 
political event contracts, where arbitrage deviations persist until liquidity deepens sufficiently to support larger executions. Our results suggest that in sports markets, where contract lifespans are measured in hours rather than weeks, this liquidity threshold is never reached, permanently confining risk-free extraction to the retail scale.

\section{Limitations}

Two constraints bound the scope and generalizability of our findings.

\textbf{Polling Latency.} The 3.6 to 5.5 second API polling cadence means 
that arbitrage episodes resolving entirely within a single polling interval 
evade detection. The frequencies reported therefore represent a conservative 
lower bound on true market inefficiency, and the measured durations an upper 
bound on actual lifespans.

\textbf{Sample Period and Market Scope.} The dataset spans a single month 
(February 4 to March 4, 2026) and is restricted to NBA game markets. Whether 
these findings generalize to other sports, longer time horizons, or other 
Polymarket domains, such as political or macroeconomic contracts with 
materially different liquidity profiles and contract durations, remains 
an open empirical question.

\section{Conclusion}

This paper presents a systematic empirical analysis of algorithmic arbitrage 
in Polymarket's NBA game markets, leveraging over 75 million high-frequency limit order book snapshots across 173 games. Our findings paint a consistent picture of a market that is structurally efficient yet not frictionless.

Single-market arbitrage is exceedingly rare: only 7 valid in-game episodes 
were identified across 3,042 markets, persisting for a median of just 3.6 
seconds before automated participants restored efficiency. Combinatorial 
arbitrage across Moneyline--Spread pairs is more frequent, yielding 290 active episodes concentrated in the final minutes of live play, where abrupt scoring events cause rapid cross-market dislocations. Yet both strategies share a common structural ceiling: shallow order book depth caps the 
exploitable size of every identified opportunity, with 76.9\% of combinatorial episodes constrained to an average executable size of just 14.8 shares.

These results ground a well-established theoretical prediction in a new 
empirical setting. The limits-to-arbitrage framework of \citet{shleifer1997limits} posits that execution frictions prevent the full correction of mispricings even when they are unambiguously identified; our 
data confirm that liquidity shallowness is precisely this friction in decentralized prediction markets. Taken together, the evidence suggests that while Polymarket NBA markets are nearly efficient in aggregate, the residual inefficiencies that do exist are not accessible at institutional scale, they are structurally confined to the retail tier, offering diminishing returns to any participant seeking to deploy meaningful capital.

\bibliographystyle{plainnat}
\bibliography{arbitrage}

\appendix

\section{Single Market Arbitrage: Data Engineering and Methodology}
\label{sec:appendix_single_market}

\subsection{Data Collection Architecture and Polling Latency}
Orderbook snapshots were collected via the Polymarket CLOB API (\texttt{GET /book} endpoint). Requests were batched by event slug to ensure high synchronization between the two sides of a binary market. Network round-trip time averaged 171 milliseconds, and server-side timestamps confirmed that competing outcomes within the same market were recorded with a discrepancy of roughly 2 to 50 milliseconds. To minimize redundant storage, updates were persisted only when the orderbook's cryptographic hash changed.

Iterating sequentially through all active NBA markets introduced a mechanical polling loop with a median interval of 3.6 to 5.5 seconds per market. Because market states were observed at these discrete intervals, the reported frequencies represent a conservative lower bound of true arbitrage occurrences, while the recorded durations serve as an upper bound for their actual lifespans.

\subsection{State Alignment and Micro-Desynchronization Mitigation}
Database inserts were restricted to actual changes in the orderbook state. If the hash for Team A changed but Team B remained identical, Team B was omitted from the API response to save bandwidth. To reconstruct the continuous market state, the last-known valid limit orders were mathematically projected forward in time. 

While the off-chain matching engine updates mirrored tokens synchronously, the 2 to 50 millisecond serialization delay in the CLOB API introduced a vulnerability. Projecting staggered timestamps forward momentarily crossed the new price of one outcome against the stale price of its complement, falsely registering sub-second phantom arbitrage. To prevent this, a dynamic time-based clustering algorithm was applied. Sequential updates occurring within 500 milliseconds of each other were classified as originating from the same API polling batch and were algorithmically synchronized to share a single discrete timestamp prior to state alignment.

\subsection{Signal Validation and Execution Constraints}
\label{sec:appendix_signal_validation}

To ensure that identified inefficiencies represent actionable trading opportunities rather than phantom signals, four strict filters were applied to the raw detection output.

\textbf{Top-of-Book Restriction (Level 1 Data):}
The algorithm exclusively evaluates the Best Bid and Best Ask prices, ignoring deeper order book liquidity. In high-frequency prediction markets, liquidity beyond the top-of-book is subject to elevated cancellation rates; by the time a trader walks the book to execute deeper orders, the liquidity profile has often changed. Restricting the analysis to Level 1 data therefore represents the most conservative, highly executable measure of available liquidity. Furthermore, because arbitrage is defined by the crossing of the spread, if the top-of-book prices fail to satisfy the profitability threshold, deeper levels mathematically cannot.

\textbf{Minimum Liquidity Constraint:}
Opportunities were recorded only if the available size at the bottleneck price (the minimum executable shares across Outcome A and Outcome B) exceeded \$10 USDC. This threshold filters out microscopic dust orders that are economically unviable after accounting for on-chain gas fees or slippage, consistent with the approach of \cite{saguillo2025arbitrage}.

\textbf{Profit Threshold:}
Because Polymarket currently imposes zero trading fees on NBA markets, the raw mathematical threshold of $Profit > \$0.00$ was maintained, capturing the pure theoretical inefficiency of the order book prior to any external gas cost considerations.

\textbf{Exclusion of Post-Game Resolution Noise:}
Order book snapshots recorded after the physical conclusion of the sporting event were strictly excluded. Once a game ends, the market enters a transitional state awaiting official third-party oracle resolution. During this window, rational market makers withdraw their active quotes, severely hollowing out the order book. The stale limit orders that remain visible via the CLOB API represent frozen liquidity that cannot be actively exploited in practice. As quantified in Table~\ref{tab:arb_exclusion_spread}, the median bid-ask spread explodes to 7,532.65 bps post-game, more than seven times wider than during live play. Excluding this phase ensures the analysis measures only true, exploitable opportunities under genuine outcome uncertainty.

\subsection{Forward Duration Measurement}
\label{sec:appendix_duration}
To measure the duration of a market state, a forward-looking difference 
calculation was utilized:
$$\text{Duration}_n = t_{n+1} - t_n$$
To ensure that exchange outages or halted markets were not mistakenly 
classified as extended arbitrage opportunities, a dynamic ``Trust Ceiling'' 
$C_{phase}$ was applied, capping the credited duration:
$$\text{Valid\_Duration}_n = \min(t_{n+1} - t_n,\ C_{phase})$$
The ceiling was defined by game phase: $C_{phase} = 1800$ seconds for 
Pre-Game and Post-Game, and $C_{phase} = 300$ seconds for In-Game, 
accounting for extended television timeouts or official reviews.

For terminal states (the final row of an episode, where $t_{n+1}$ does not 
exist), the preprocessing script calculated the exact median gap between all 
unique snapshots for that specific market. This dynamically calculated 
latency (typically $\sim\!4.0$ seconds) served as the mathematical baseline, 
ensuring terminal events were credited precisely with the natural mechanical 
heartbeat of the data collector.

\subsection{The Transition from Snapshots to Episodes}
Consecutive arbitrage snapshots were grouped into distinct continuous market events, or episodes. Crucially, the profitability of an episode was calculated using a ``One-Shot'' execution paradigm. Summing the profit of multiple snapshots within a single episode would falsely assume regenerating liquidity and drastically inflate theoretical returns. Instead, the analysis script evaluated all snapshots within an episode's time boundaries and extracted the single maximum realizable profit.

\section{Combinatorial Arbitrage: Data Engineering and Methodology}
\label{sec:appendix_combinatorial}

\subsection{Signal Validation and Execution Constraints}
\label{sec:appendix_combinatorial_signal_validation}

To ensure that identified combinatorial inefficiencies represent actionable
trading opportunities, the same execution filters applied in
Appendix~\ref{sec:appendix_signal_validation} were enforced, alongside a
combinatorial-specific execution condition.

\textbf{Combinatorial Execution Condition:}
Because Polymarket does not support direct short selling, the strategy
constructs a synthetic short by purchasing the complement of the overpriced
Spread token (see Section~\ref{sec:problem_setting}). The risk-free execution
condition requires the combined ask prices of both legs to be strictly less
than the guaranteed \$1.00 payout:
$$\text{Ask}(ML_A) + \text{Ask}(Sp_B) < 1.00$$
Only states satisfying this strict inequality were flagged as candidate
arbitrage episodes.

\textbf{Top-of-Book Restriction (Level 1 Data):}
Identical to the single-market analysis, the algorithm exclusively evaluates
the Best Ask prices on both legs. If the top-of-book combined cost fails to
satisfy the execution condition, deeper levels mathematically cannot, making
the Level 1 restriction both conservative and exhaustive.

\textbf{Minimum Liquidity Constraint:}
Opportunities were recorded only if the available size at the bottleneck leg, the leg with the lower resting volume, exceeded \$10 USDC, filtering
out dust orders that are economically unviable in practice.

\textbf{Profit Threshold:}
Given that Polymarket imposes zero trading fees on NBA markets, the raw
threshold of $Profit > \$0.00$ was maintained, capturing the full theoretical
inefficiency of the combined order book state.

\textbf{Exclusion of Post-Game Resolution Noise:}
As established in Appendix~\ref{sec:appendix_signal_validation}, all combinatorial states recorded after the physical conclusion of the game were
excluded.

\subsection{Generation of the Unified Event Timeline}
Detecting combinatorial arbitrage between two dependent prediction markets introduces a synchronization challenge. Because Polymarket's order books for distinct markets update asynchronously, Market A and Market B rarely register state changes at the exact same millisecond.

To accurately reconstruct the combinatorial market state at any microsecond, a Unified State Machine approach was implemented. By performing a full outer join on the UTC timestamps of the Moneyline ($ML$) and Spread ($S$) updates, we created a master sequence of events. Let $T_{ML}$ be the set of timestamps where the Moneyline market updated, and $T_{S}$ be the set where the Spread market updated. The unified timeline $T_U$ is defined as:
$$T_U = T_{ML} \cup T_{S}$$
At any index $n$ in this timeline ($t_n \in T_U$), an update occurs in at least one of the two markets, defining a new potential combinatorial state.

\subsection{State Imputation} 
Because an update at $t_n$ typically involves only one market, the state of the other market must be imputed to evaluate the combinatorial price. To avoid look-ahead bias, a strict forward-fill operation was applied. The imputed price for a market $M$ at time $t_n$ is the price from its most recent recorded update:
$$P_M(t_n) = P_M(t_i) \quad \text{where} \quad t_i = \max \{t \in T_M \mid t \le t_n\}$$
This guarantees that at any timestamp $t_n$, the algorithm is only utilizing the limit orders that were verifiably resting on the exchange at that exact moment.

\subsection{Combinatorial Duration and Episode Measurement}
Duration measurement follows the same forward-looking difference framework 
and phase-dependent Trust Ceiling $C_{phase}$ defined in 
Appendix~\ref{sec:appendix_duration}. The unified timeline $T_U$ introduces 
one additional consideration: because state changes now originate from two 
asynchronous order books, the forward gap $t_{n+1} - t_n$ can be 
substantially shorter than in the single-market case, reflecting updates 
from either the Moneyline or Spread book. The trust ceiling and terminal 
state fallback ($\approx\!4.0$ seconds) are applied identically, ensuring 
that combinatorial profit calculations represent strictly executable temporal 
windows.

\end{document}